# Kinematic Moment of Inertia of e-e Rare Earths Nuclei

**Mohamed E. Kelabi**
Physics Department, University of Tripoli, Tripoli, LIBYA

**Abstract**

The kinematic moment of inertia of the rare earth even-even nuclei was calculated using three parametric energy based expression. The plot of kinematic moment of inertia versus nuclear spin shows a better sensitivity to backbending than energy plot.

**Introduction**

It is well know that the moment of inertia is a measure of the resistance of a body to change its rotation. The classical expression for the rotational energy is given by

$$E = \frac{1}{2\mathcal{J}} L^2 \qquad (1)$$

which can be transformed into a quantized expression. Using quantum mechanics aspects for rigid rotor [1], [2], one easily obtains

$$E = \frac{\hbar^2}{2\mathcal{J}} I(I+1). \qquad (2)$$

This simple form is used to study the ground state rotational band of even-even nuclei, where the total angular momentum (nuclear spin) can take the sequence $I = 0, 2, 4, \ldots$ and even parity, while the nuclear moment of inertia $\mathcal{J}$ is found to increase with angular momentum [3], [4]. It is thereafter known as kinematic moment of inertia $\mathcal{J}^{(1)}$, and generally is given by

$$\frac{\mathcal{J}^{(1)}}{\hbar^2} = \frac{1}{2} \left( \frac{dE}{d\,I(I+1)} \right)^{-1}. \qquad (3)$$

This expression can be simplified to the following experimental form

$$\frac{\mathcal{J}^{(1)}}{\hbar^2} = \frac{2I-1}{E_\gamma} \qquad (4)$$

$$E_\gamma = E(I) - E(I-2) \qquad (5)$$

At relatively high spin states Eq. (2) shows an underestimate values of energy levels, compared with experiment [5]. To compensate for such apparent discrepancies many attempts were created [6], [7], [8], [9], [10]. An extended form of Eq. (2) was manifested by adding a correction term [11], of the form

$$E = \frac{\hbar^2}{2\mathcal{J}} I(I+1) + B\, I^2(I+1)^2 . \qquad (6)$$

**Formalisms**

As proposed in [12], we use two parametric nuclear softness expression (NS2) [8],

$$E_I = \frac{A\, I(I+1)}{1+\sigma_1 I}, \qquad (7)$$

where the softness parameter [13],

$$\sigma_1 = \frac{1}{\mathcal{J}_0} \left.\frac{\partial \mathcal{J}(I)}{\partial I}\right|_{I=0}, \qquad (8)$$

and merge into Eq. (6), one obtains

$$E = \frac{A\, I(I+1)}{1+\sigma_1 I} + B\, I^2(I+1)^2 \qquad (9)$$

where 
$$A = \frac{\hbar^2}{2\mathcal{J}_0} .$$

In this work, we solve Eq. (9) by fitting for the three unknowns: $A$, $\sigma_1$, and $B$, then using Eq. (4) to extract the kinematic moment of inertia $\mathcal{J}^{(1)}$.

**Results**

We present in Fig. 1 the results of energy fits using Eq. (9). In Fig. 2 we show the calculated kinematic moment of inertia using Eq. (4). The calculations are based on data taken from the references [14], for $^{150}$Sm, and [15], for all other nuclei. A list of fitting parameters is given in Table 1.

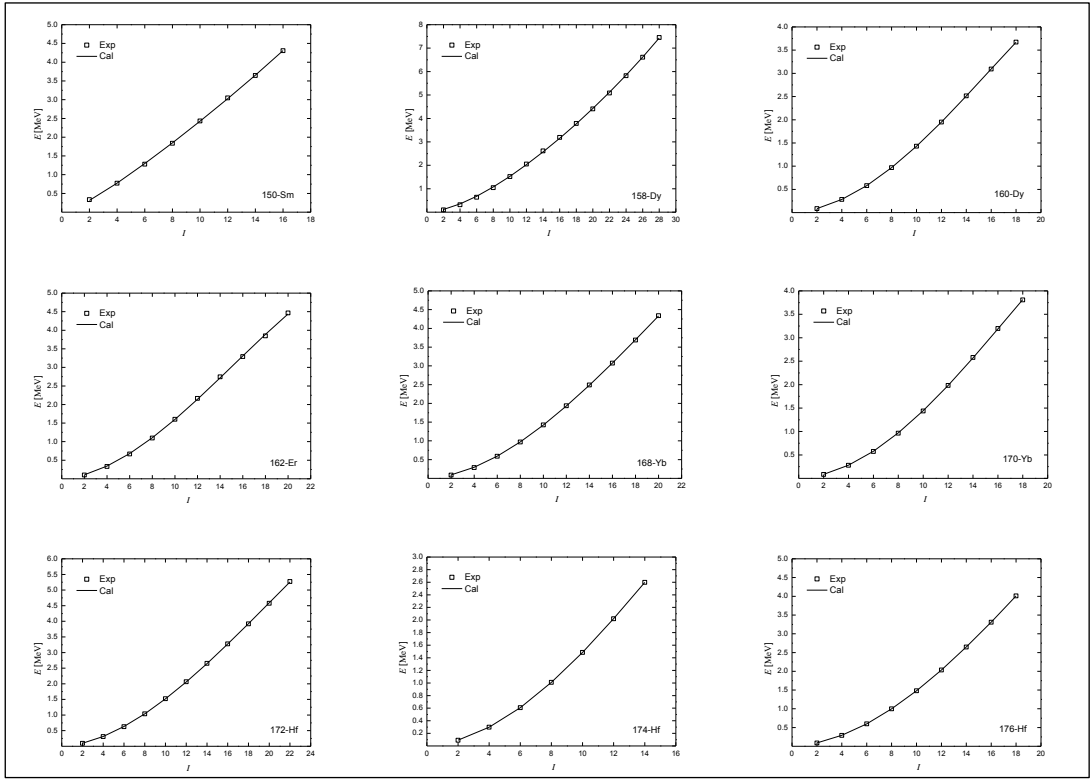

Figure 1. Plots of energy versus angular momentum.

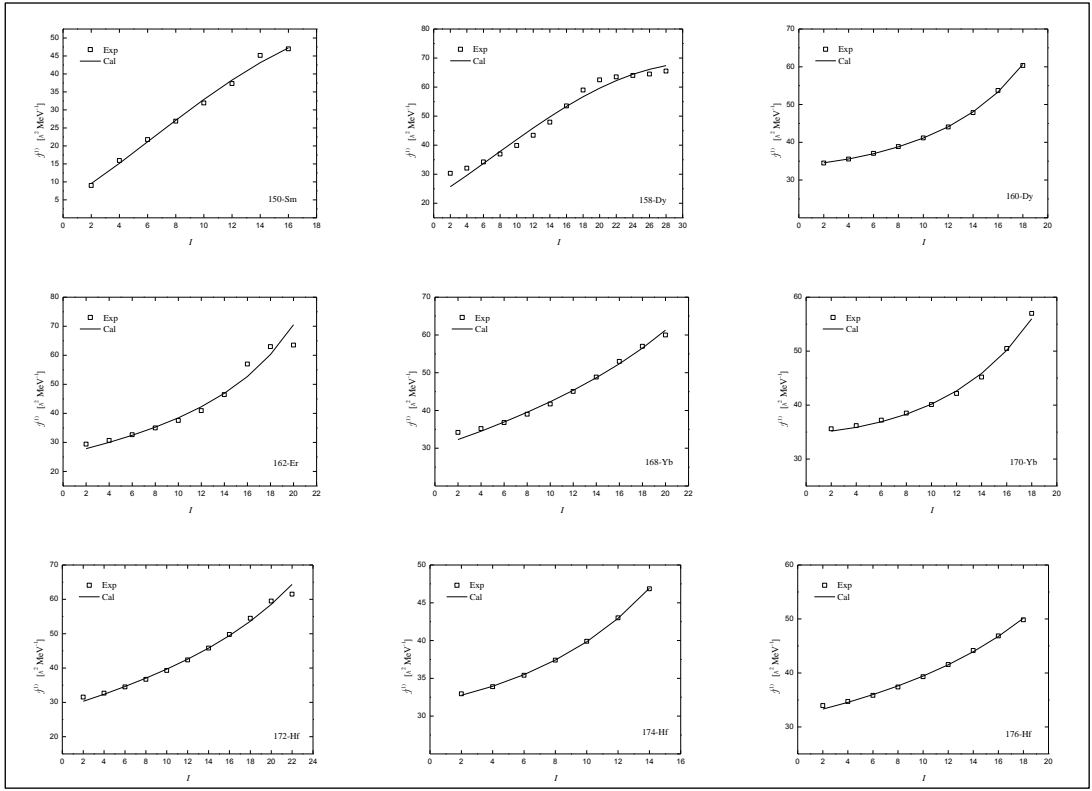

Figure 2. Plots of kinematic moment of inertia versus angular momentum.

| Nucleus | $A$ [MeV] | $\sigma_1$ [$1/\hbar$] | $B \times 10^{-6}$ [MeV] | $\mathcal{J}_0 = 1/2A$ [$\hbar^2$/MeV] |
|---|---|---|---|---|
| 150-Sm | 0.08061 | 0.27008 | -2.54404 | 6.202704 |
| 158-Dy | 0.02166 | 0.05713 | -1.00986 | 23.08403 |
| 160-Dy | 0.01500 | 0.00709 | 6.76900 | 33.33333 |
| 162-Er | 0.01889 | 0.02556 | 4.58236 | 26.46903 |
| 168-Yb | 0.01623 | 0.02390 | 1.59122 | 30.80715 |
| 170-Yb | 0.01435 | 0.00339 | 6.92966 | 34.84321 |
| 172-Hf | 0.01727 | 0.02326 | 2.03506 | 28.95194 |
| 174-Hf | 0.01559 | 0.00959 | 6.58017 | 32.07184 |
| 176-Hf | 0.01539 | 0.01183 | 2.81541 | 32.48863 |

Table 1. Fitting parameters. The unperturbed moment of inertia is given on the right column.

**Conclusion**

From the above given Figures, the calculated moment of inertia shows fairly good results compared with experiment. The unperturbed moment of inertia, corresponds to ground state band, is obtained. We noted some deviation in our results especially for nuclei which possess backbending phenomenon. Much improvement in the results can be achieved with more parameters instead of only three. The plot of kinematic moment of inertia versus angular momentum shows a noticeable sensitivity for backbending phenomena compared with the plot of energy versus moment of inertia.